\RequirePackage{fix-cm}
\documentclass[smallextended]{svjour3}       % onecolumn (second format)
\smartqed  % flush right qed marks, e.g. at end of proof
\usepackage{graphicx}
\usepackage[utf8]{inputenc}

\usepackage{amsfonts}
\usepackage{mathtools}
\usepackage{braket}
\usepackage{slashed}
\usepackage{color}
\usepackage{amsmath}
\usepackage{esint}
\usepackage{amssymb}
\usepackage{bbold}

\newcommand{\be}{\begin{equation}}
	\newcommand{\ee}{\end{equation}}
\newcommand{\bea}{\begin{eqnarray}}
	\newcommand{\eea}{\end{eqnarray}}

\newcommand{\p}{\partial}

\def\cL{{\cal L}}
\def\cM{{\cal M}}

\def\cR{{\cal R}}

\newcommand{\gb}{\bar{g}}
\newcommand{\Nb}{\bar{N}}
\newcommand{\Nh}{\hat{N}}
\newcommand{\sib}{\bar{\sigma}}
\newcommand{\sh}{\hat{\sigma}}
\newcommand{\Db}{\bar{D}}
\newcommand{\Kb}{\bar{K}}
\newcommand{\Rb}{\bar{R}}

\begin{document}

\title{Functional Renormalization Group flows on  \\[1ex] Friedman-Lema\^{\i}tre-Robertson-Walker backgrounds \thanks{Proceedings based on the talk given by F.S.\ at  the workshop Lema\^{\i}tre, ``Black Holes, Spacetime Singularities and Gravitational Waves'', held at the Vatican observatory, May 9$^{\rm th}$ - 12$^{\rm th}$ 2017.}
}

\titlerunning{Functional Renormalization Group flows on FLRW backgrounds}

\author{Alessia Platania        \and Frank Saueressig
         %etc.
}

\institute{ Department of Physics and Astronomy, University of Catania, \\ Via S. Sofia 63, 95123 Catania, Italy\\ INFN, Catania Section, via S. Sofia 64, 95123 Catania, Italy \\ INAF, Catania Astrophysical Observatory, via S. Sofia 78, 95123 Catania, Italy \\
	\email{alessia.platania@oact.inaf.it}  \\          
           \and
           Institute for Mathematics, Astrophysics and Particle Physics (IMAPP),\\
           Radboud University Nijmegen, Heyendaalseweg 135, 6525 AJ Nijmegen, The Netherlands
           \email{f.saueressig@science.ru.nl} \\
}

\date{}

\maketitle

\begin{abstract}
We reanalyze the construction of the gravitational functional renormalization group equation tailored to the Arnowitt-Deser-Misner formulation emphasizing its connection to the covariant formulation. The results obtained from projecting the renormalization group flow onto the Einstein-Hilbert action  are reviewed in detail and we provide a novel example illustrating how the formalism may be connected to the Causal Dynamical Triangulations approach to quantum gravity.
\end{abstract}

%------------------------------------------------------------------
\section{Introduction}
\label{intro}
%------------------------------------------------------------------
One of the most intriguing features of classical general relativity is the occurrence of spacetime singularities in its solutions. According to the $\Lambda$CDM model, constituting the standard model underlying cosmology, the universe is born from a phase where the energy density diverges (initial singularity). Moreover, the gravitational collapse of a star may result in the formation of a black hole accommodating a spacetime singularity in its interior. These situations are characterized by energy densities or spacetime curvatures exceeding their Planck values. On this basis it is expected that a description of spacetime in these extreme regions requires a more complete framework which is broadly referred to as quantum gravity.

A very conservative proposal geared towards defining a quantum theory of the gravitational interactions covering all length scales is Asymptotic Safety
\cite{Niedermaier:2006wt,Codello:2008vh,Reuter:2012id,Roberto:book}. This approach stays within the framework of relativistic quantum field theory and builds on Wilson's modern formulation of renormalization. Its key ingredient is an interacting renormalization group (RG) fixed point which controls the gravitational interactions at transplanckian energy scales. The fixed point ensures that the theory does not suffer from unphysical ultraviolet (UV) divergences. Moreover, it may come with a predictive power similar to the one known from perturbatively renormalizable quantum field theories. In this setting general relativity arises dynamically when the RG flow undergoes a crossover from the interacting fixed point (quantum regime) to the free fixed point (classical regime) \cite{Reuter:2001ag}. Moreover, the phase diagram accommodates RG trajectories which give rise to the observed values of Newton's coupling and the cosmological constant on macroscopic scales \cite{Reuter:2004nx}.

Most investigations related to asymptotically safe gravity rely on functional renormalization group methods and encode the gravitational degrees of freedom in fluctuations of the spacetime metric $g_{\mu\nu}$ (covariant approach). In this proceedings we focus on an alternative setting based on the Arnowitt-Deser-Misner or $(3+1)$-formulation of gravity, see \cite{Gourgoulhon:2007ue} for a pedagogical introduction. This construction equips spacetime with a foliation structure by welding together spatial slices where the time-coordinate $\tau$ is constant. This introduces a preferred direction which may be interpreted as an (Euclidean) time direction and provides a causal structure. As a byproduct it gives access to interesting observables comprising, e.g.\ the expectation value  for spatial volumes $V_3(\tau)$ as a function of time. This facilitates the direct comparison with results obtained from the Causal Dynamical Triangulations program \cite{Ambjorn:2012jv} which evaluates the gravitational partition sum by Monte Carlo methods. Moreover, the close relation of the ADM framework to cosmology allows to address questions related to the very early universe based on first principles.

The proceedings is organized as follows. Sect.\ \ref{sect:2} reviews the ADM formalism and the most important elements entering into the construction of the corresponding functional renormalization group equation (FRGE). Approximate solutions of the FRGE based on the Einstein-Hilbert action are discussed in Sect.\ \ref{sect:3} and novel results relating the FRGE to the Causal Dynamical Triangulations program are presented in Sect.\ \ref{sect:4}. We close by summarizing the status and perspectives of the program in Sect.\ \ref{sect:5}.
 
%-----------------------------------------------------------------
\section{The functional renormalization group in the ADM-formalism}
\label{sect:2}
%-----------------------------------------------------------------
In this section we provide the geometrical background underlying the formulation of gravity in the Arnowitt-Deser-Misner (ADM) formalism and highlight the main properties of the resulting functional renormalization group equation (FRGE) for the effective average action.
%-----------------------------------------------------------
\subsection{Gravity in the Arnowitt-Deser-Misner formulation}
\label{sect.2a}
%-----------------------------------------------------------
The ADM formulation of gravity starts from a $D$-dimensional (Euclidean) spacetime $\cM$ with metric $g_{\mu\nu}$ carrying coordinates $x^\mu$. Subsequently, the construction introduces a time-function $\tau(x)$ which assigns a specific time to each spacetime point. The set of points with the same value of $\tau(x)$ form $d$-dimensional spatial slices $\Sigma_{\tau_i} \equiv \{x : \tau(x) = \tau_i \}$. The gradient of the time function $n_\mu \equiv N \p_\mu \tau(x)$ defines a vector $n^\mu$ normal to the spatial slices. The lapse function $N(x)$ ensures that $g_{\mu\nu} n^\mu n^\nu = 1$. One then introduces a new coordinate system $x^\mu \mapsto (\tau, y^i)$, $i=1,\ldots,d$ where the $y^i$ provide coordinates on $\Sigma_{\tau}$. Defining the vector field $t^\mu$ through the relation $t^\mu \p_\mu \tau = 1$,  the coordinate systems on neighboring spatial slices can be related by requiring that the coordinate $y^i$ is constant along the integral curves of $t^\mu$. 

The tangent space at a point in $\cM$ can then be decomposed into a subspace spanned by vectors tangent to $\Sigma_{\tau}$ and its complement.  The corresponding basis vectors are constructed from the Jacobians
\be\label{proj1}
t^\mu = \left . \frac{\partial x^\mu}{\partial \tau} \right|_{y^i} \, , \qquad e_i{}^\mu = \left. \frac{\p x^\mu}{\p y^i} \right|_\tau \, . 
\ee
The normal vector thus satisfies $g_{\mu\nu} \, n^\mu \, e_i{}^\nu = 0$ and the metric induced on the spatial slices is given by
\be
\sigma_{ij}(\tau,y) \equiv e_i{}^\mu \, e_j{}^\nu \, g_{\mu\nu} \, . 
\ee   
A priori $t^\mu$ is neither tangent nor normal to the spatial slices. The Jacobians \eqref{proj1} imply that its decomposition into components normal and tangent to $\Sigma$ is given by 
\be\label{tdec}
t^\mu = N \, n^\mu + N^i \, e_i{}^\mu \, ,
\ee
where $N^i(\tau,y)$ is called the shift vector. Furthermore, eq.\ \eqref{proj1} implies that the coordinate one-forms in the two coordinate systems are related by
\be\label{t2}
dx^\mu = t^\mu d\tau + e_i{}^\mu dy^i = N n^\mu d\tau + e_i{}^\mu \, (dy^i + N^i d\tau) \, . 
\ee
Combining the relations \eqref{t2} and \eqref{tdec} with the normal property of $n^\mu$ and the definition of $\sigma_{ij}$, the line-element $ds^2 = g_{\mu\nu} dx^\mu dx^\nu$ can be recast in terms of the ADM-fields $\{N, N_i, \sigma_{ij}\}$:
\be\label{fol1}
ds^2 
=  N^2 d\tau^2 +  \sigma_{ij} \, (dy^i + N^i d \tau) 
(dy^j +  N^j d \tau) \, . 
\ee 
For the components of the metric tensor the decomposition \eqref{fol1} implies the following relations
\be\label{metcomp}
g_{\mu\nu} = \left(
\begin{array}{cc}
	N^2 + N_i N^i \; \;  & \;  \; N_j \\
	N_i &  \sigma_{ij} 
\end{array}
\right) \, , \qquad 
g^{\mu\nu} = \left(
\begin{array}{cc}
	\frac{1}{N^{2}} \; \;  & \;  \; -  \frac{N^j}{ N^{2}}  \\
	-  \frac{N^i}{ N^{2}}  	 \; \;  & \;  \;  \sigma^{ij} +  \, \frac{ N^i \,  N^j}{ N^{2}} 
\end{array}
\right) \,
\ee
where spatial indices $i,j$ are raised and lowered with $\sigma_{ij}$. This entails that the relation between $g_{\mu\nu}$ and the ADM fields is actually non-linear. 

An infinitesimal coordinate transformation $v^\mu(\tau,y)$ acting on the metric can be expressed in terms of the Lie derivative $\cL_v$
\be\label{diffeo1}
\delta \gamma_{\mu\nu} = \cL_v \, \gamma_{\mu\nu} \, . 
\ee 
Decomposing $v^\mu = \left(f(\tau,y), \zeta^i(\tau,y)\right)$ 
into its temporal and spatial parts, the transformation \eqref{diffeo1} determines the transformation properties of the component fields under Diff($\cM$)
\be\label{eq:gaugeVariations}
\begin{split}
	\delta N &= \p_\tau (f N ) + \zeta^k \p_k N  - N N^i\p_i f \, , \\
	\delta  N_i &= \partial_\tau( N_i f) + \zeta^k\partial_k  N_i +  N_k\partial_i\zeta^k
	+ \sigma_{ki}\partial_\tau \zeta^k
	+  N_k  N^k\partial_i f  +  N^2\partial_i f \, , \\
	\delta\sigma_{ij} &= f\p_\tau \sigma_{ij} + \zeta^k\p_k \sigma_{ij} + \sigma_{jk}\p_i\zeta^k + \sigma_{ik}\p_j\zeta^k + N_j\p_i f + N_i\p_j f  \, . 
\end{split}
\ee
Thus, at the level of the component fields the action of Diff($\cM$) is non-linear. Notably, Diff($\cM$) contains the subgroup of foliation preserving diffeomorphisms Diff($\cM$, $\Sigma$) where $f = f(\tau)$ is restricted to be independent of the spatial coordinates.  Inspecting \eqref{eq:gaugeVariations}, one observes that the restriction $f(\tau,y) \rightarrow f(\tau)$ eliminates the non-linear terms from the transformation laws so that the component fields transform linearly with respect to this  subgroup. 

%-----------------------------------------------------------
\subsection{The functional renormalization group equation on foliated spacetime}
%-----------------------------------------------------------
The primary tool for investigating gravitational RG flows at a non-perturbative level is the FRGE for the effective average action $\Gamma_k$  \cite{Wetterich:1992yh,Morris:1993qb,Reuter:1993kw,Reuter:1996cp}. In the context of gravity, its construction hinges on the background field formalism which decomposes the physical metric $g_{\mu\nu}$ into a fixed but arbitrary reference background $\gb_{\mu\nu}$ and fluctuations $h_{\mu\nu}$ around this background. Prominent choices for this decomposition are the linear split
\be\label{linsplit}
g_{\mu\nu} = \gb_{\mu\nu} + h_{\mu\nu}
\ee
and the exponential split
\be\label{expsplit}
g_{\mu\nu} = \gb_{\mu\alpha} \, \left[ e^h \right]^\alpha{}_\nu \, . 
\ee
The virtue of the background field method is that the quantum theory inherits a symmetry under background field transformations where any field transforms as a tensor of the corresponding rank. At the same time $\gb_{\mu\nu}$ provides a reference scale which allows to construct a RG scale $k$. It is expected that the two decompositions \eqref{linsplit} and \eqref{expsplit} lead to different quantum theories, see \cite{Nink:2015lmq} for an explicit demonstration in two spacetime dimensions. Heuristically, this may be understood as follows. Fixing the background $\gb_{\mu\nu}$, the fluctuations in the linear split may change the signature of $g_{\mu\nu}$ while the exponential split guarantees that both $g_{\mu\nu}$ and $\gb_{\mu\nu}$ have the same signature. Thus the set of fluctuations considered in the two settings actually differs. At the level of the path-integral this difference is reflected by different choices for the path-integral measure.
 
In the ADM formalism the gravitational degrees of freedom are carried by $N$, $N_i$, and $\sigma_{ij}$. Following the background field approach used in the covariant framework, these fields are decomposed into fixed but arbitrary background values (marked with bars) and fluctuations (marked with hats). The linear ADM split then uses
\be\label{linearsplit}
N = \Nb + \Nh \, , \qquad N_i = \Nb_i + \Nh_i \, , \qquad \sigma_{ij} = \sib_{ij} + \sh_{ij} \, .
\ee 
For convenience, we denote the collection of physical fields, background fields, and the fluctuations by $\chi$, $\bar{\chi}$, and $\hat{\chi}$, respectively and set $\hat{\sigma} \equiv \sib^{ij} \hat{\sigma}_{ij}$.
Similarly to \eqref{expsplit}, the linear split of $\sigma_{ij}$ can be replaced by $\sigma_{ij} = \sib_{ik} \left[ e^{\hat{\sigma}} \right]^k{}_j$, defining the  exponential ADM split. The properties of these choices are conveniently discussed by writing the determinant of the spacetime metric in terms of the ADM fields
\be
\det g = N^2 \, \det \sigma \, . 
\ee
Combining this relation with the decomposition \eqref{linearsplit} one concludes that the fluctuations can not change the signature in the (Euclidean) time direction. The exponential ADM split furthermore fixes the signature of the metric on the spatial slices. Thus the exponential ADM split has similar properties as the exponential split in the covariant setting. 

When setting up the FRGE in the ADM formalism, it is actually useful that there is a local map relating the fluctuations of the covariant and the ADM formulation. This allows to understand the ADM construction as a particular, background-dependent redefinition of the fluctuation fields. The explicit relation between the two settings is found by starting from the linear split \eqref{linsplit} and performing an ADM decomposition of both $g_{\mu\nu}$ and $\gb_{\mu\nu}$ according to \eqref{metcomp}. The result allows to express the components of $h_{\mu\nu}$ in terms of $\Nh, \Nh_i$, and $\hat{\sigma}_{ij}$ and their background values
\be\label{map1}
\begin{split}
	h_{00} = & \, 2 \Nb \Nh + \Nh^2 + \sigma^{ij} (\Nb_i + \Nh_i)(\Nb_j + \Nh_j) - \sib^{ij} \Nb_i \Nb_j \, , \\
	h_{0i} = & \, \Nh_i \, , \\
	h_{ij} = & \, \hat{\sigma}_{ij} \, . 
\end{split}
\ee
Note that the resulting map is actually non-linear. Owed to the presence of $\sigma^{ij}$ in $h_{00}$ it involves the spatial fluctuations $\hat{\sigma}_{ij}$ to arbitrary high orders. At the linear level the map \eqref{map1} entails
\be\label{map1lin}
h_{00} \approx 2 \Nb \Nh - \hat{\sigma}^{ij} \Nb_i \Nb_j + 2 \, \sib^{ij} \Nh_i \Nb_j \, , \qquad
h_{0i} = \Nh_i \, , \qquad
h_{ij} = \hat{\sigma}_{ij} \, .
\ee

The construction of the FRGE for the effective average action $\Gamma_k$  then proceeds along the same lines as in the covariant case \cite{Reuter:1996cp} and gives \cite{Rechenberger:2012dt}
\be\label{FRGE}
 \p_t \Gamma_k[\hat{\chi};\bar{\chi}] = \frac{1}{2} \, {\rm STr} \left[ \left( \Gamma_k^{(2)} + \cR_k \right)^{-1} \,  \p_t \cR_k \right] \, . 
\ee
Here $t \equiv \ln(k/k_0)$ denotes the RG time, $\Gamma_k^{(2)}$ is the second functional derivative of $\Gamma_k$ with respect to the fluctuation fields $\hat{\chi}$, and the ${\rm STr}$ contains an integral over loop-momenta as well as a sum over component fields. The regulator $\cR_k$ provides a $k$-dependent mass term for fluctuations with momenta $p^2 \lesssim k^2$ and vanishes if $p^2 \gg k^2$. The interplay of $\cR_k$ appearing in the numerator and denominator ensures that the flow of $\Gamma_k$ is driven by quantum fluctuations with momentum $p^2 \approx k^2$. In this way $\Gamma_k$ provides a one-parameter family of effective actions encoding the dynamics of a physical system at the scale $k$. In connection with the exponential ADM split the flow equation \eqref{FRGE} provides a complementary tool for probing the spacetime structure seen in Monte Carlo simulations carried out within the CDT program \cite{Ambjorn:2012jv}.

At this stage the following remark is in order. A key ingredient in the construction of the FRGE is the property that the $k$-dependent mass term giving rise to the regulator $\cR_k$ is quadratic in the fluctuation fields. At the same time eq.\ \eqref{eq:gaugeVariations} indicates that the ADM fields transform non-linearly under diffeomorphisms. The FRGE then realizes only the linear part of the symmetry group as a background symmetry. Thus \eqref{FRGE} is invariant under foliation preserving background transformations only.

%-----------------------------------------------------------
\section{Renormalization group flows in the Einstein-Hilbert truncation}
\label{sect:3}
%-----------------------------------------------------------
We illustrate the working of the FRGE by approximating the gravitational part of $\Gamma_k$ by the Einstein-Hilbert action. In terms of the ADM fields one then has
\be\label{GammaEH}
\Gamma_k^{\rm grav} \simeq \frac{1}{16 \pi G_k} \int d\tau d^dy \, N \sqrt{\sigma} \left[ K_{ij} K^{ij} - K^2 - R + 2 \Lambda_k \right] \, ,
\ee
where the extrinsic curvature is defined as
\be\label{Kext}
K_{ij} \equiv \frac{1}{2 N} \left( \p_\tau \sigma_{ij} - D_i N_j - D_j N_i \right) \, , \quad K \equiv \sigma^{ij} K_{ij} \, , 
\ee
and $R$ and $D_i$ are the Ricci scalar and covariant derivative constructed from $\sigma_{ij}$. The ansatz comprises two scale-dependent couplings, the cosmological constant $\Lambda_k$ and Newton's coupling $G_k$. The beta functions encoding the scale-dependence of these couplings may be obtained by substituting the ansatz into the FRGE and extracting suitable geometric terms on both sides. Eq.\ \eqref{GammaEH} suggests evaluating the flow equation at the background level, i.e.\ setting $\hat{\chi} = 0$ after taking the necessary variations. In this case, it then suffices to keep track of terms which are at most quadratic in the fluctuation fields as these are necessary to construct the Hessian $\Gamma_k^{(2)}$. By investigating the structure of the resulting equation, it then turns out that the computation can be simplified by using a flat Euclidean FLRW background where
\be\label{FLRWbackground}
\Nb = 1 \, , \qquad \Nb_i = 0 \, , \qquad \sib_{ij}(\tau) = a(\tau)^2 \, \delta_{ij} \, , \qquad \Db_i = \p_i \, .  
\ee
At the geometric level this entails the identities
\be
\bar{K}_{ij} = \tfrac{1}{d} \sib_{ij} \bar{K} \, , \qquad \Rb = 0 \, . 
\ee
The scale-dependence of the cosmological constant and Newton's coupling can be read off from the terms proportional to the background volume and the integrated squared external curvature terms, respectively.

A key obstacle in the actual construction of flows in the ADM formalism results from the lapse $N$ and shift $N_i$ which enter \eqref{GammaEH} as Lagrange multipliers. At the level of the FRGE, which is based on an off-shell formalism, this leads to a Hessian $\Gamma_k^{(2)}$ which is degenerate. Imposing proper-time gauge which fixes $N$ and $N_i$ to their background values does not resolve this problem. An interesting route for addressing this obstacle efficiently is to start from the background gauge-fixing procedure used in the covariant formulation
\be
\Gamma_k^{\rm gf} = \frac{1}{32 \pi G_k \alpha} \int d^Dx \sqrt{\gb} \, \gb^{\mu\nu} F_\mu F_\nu \, .
\ee
Here the one-parameter family of gauge conditions $F_\mu$ are linear in the fluctuation field $h_{\mu\nu}$
\be\label{Fmu}
F_\mu = \Db^\nu h_{\mu\nu} - \beta \, \Db_\mu h \, , 
\ee
where $h \equiv \gb^{\mu\nu} h_{\mu\nu}$, and $\alpha, \beta$ are free gauge-parameters. The harmonic gauge is obtained by setting $\alpha = 1, \beta = \tfrac{1}{2}$ while the geometric gauge sets $\beta = 1/D$ and takes the limit $\alpha \rightarrow 0$. By construction it is clear that all gauge-fixing terms in this class preserve background diffeomorphism invariance. The analogous gauge conditions in the ADM formalism are obtained by substituting the map \eqref{map1} into eq.\ \eqref{Fmu}. At the linear level and for the specific background \eqref{FLRWbackground} this results in
 \be\label{gfgeneral}
 \begin{split}
 F = & \, 2 \Db_\tau \Nh + \p_i \Nh^i - \beta \, \Db_\tau \left( 2 \Nh + \hat{\sigma} \right) \, , \\
 F_i = & \, \Db_\tau \Nh_i + \p^j \hat{\sigma}_{ij} - \beta \p_i \left( 2 \Nh + \hat{\sigma} \right) \, , 
 \end{split}
 \ee
where $F_\mu = (F,F_i)$ has been decomposed in a spatial and time-component. The ghost action for this class of gauge-fixings can then be obtained in a standard way.

At this stage it is instructive to combine the gravitational and gauge-fixing terms and write down the part of the action quadratic in the fluctuation fields on flat space, $\Kb = 0$.  Abbreviating $\Delta \equiv - \sib^{ij} \p_i \p_j$ and setting $\alpha = 1$ this results in
\be\label{hessian:flat}
\begin{split}
& \, \left(32 \pi G_k \right) \left( \tfrac{1}{2} \, \delta^2 \Gamma_k^{\rm grav} + \Gamma^{\rm gf}_k \right) = \\
& \; \int  d\tau d^dy \, \sqrt{\sib} \, \Big\{ \tfrac{1}{2} \hat{\sigma}_{ij} \left[ - \p_\tau^2 + \Delta - 2 \Lambda_k  \right] \hat{\sigma}^{ij}
 + \Nh^i \left[ - \p_\tau^2 + \Delta\right] \Nh_i \\
 & \qquad \qquad \quad - \tfrac{1}{2} \hat{\sigma} \left[ (1-2 \beta^2) (- \p_\tau^2 + \Delta) - \Lambda_k \right] \hat{\sigma} \\
 & \qquad \qquad \quad + 4 \Nh \left[ (1-\beta)^2 (-\p_\tau^2) + \beta^2 \Delta \right] \Nh \\
 & \qquad \qquad \quad - 2 \Nh \left[ 2 \beta (1-\beta) (-\p_\tau^2) + (1-2 \beta^2) \Delta - \Lambda_k \right] \hat{\sigma} \\
 & \qquad \qquad \quad - (1-2\beta) (2 \Nh + \hat{\sigma}) \left( \p_i \p_j \hat{\sigma}^{ij} + 2 \p_\tau \p_i \Nh^i \right)
\Big\} \, . 
\end{split}
\ee
This shows that there is a \emph{unique gauge choice}, corresponding to harmonic gauge $\beta = 1/2$, where all derivatives combine into the $D$-dimensional Laplacian on flat space $-\p_\tau^2 + \Delta$.\footnote{Other interesting gauge-choices are of Landau-type sending $\alpha \rightarrow 0$. Moreover, taking the limit $\beta \rightarrow \pm \infty$ will gauge-fix the combination $(2\Nh + \sh)$ sharply. For an instructive discussion at the level of the covariant theory see \cite{Ohta:2016npm,Ohta:2016jvw}.} At the same time, the result demonstrates that there is no (local) gauge choice which preserves background diffeomorphism invariance and just provides a partial gauge-fixing for the lapse $\Nh$.

The connection of \eqref{hessian:flat} to cosmology is made by an additional field decomposition
\be\label{fielddec}
\begin{split}
\Nh_i = & u_i + \p_i \frac{1}{\sqrt{\Delta}} B \, , \\
\sh_{ij} = & h_{ij} - \left( \sib_{ij} + \p_i \p_j \, \tfrac{1}{\Delta} \right)  \psi + \p_i \p_j \, \tfrac{1}{\Delta} \, E + \p_i \tfrac{1}{\sqrt{\Delta}} v_j + \p_j \, \tfrac{1}{\sqrt{\Delta}} \, v_i \, ,  
\end{split}
\ee
where the component fields are subject to the constraints
\be
\partial^i u_i = 0 \, , \qquad
\p^i \, h_{ij} = 0 \, , \quad \sib^{ij} h_{ij} = 0 \, , \quad \p^i v_i = 0 
 \, , \qquad \sh \equiv \sib^{ij} \sh_{ij} \, .
\ee
In this way the FRGE on a FLRW background may directly be formulated in terms of the fields commonly used in cosmic perturbation theory.

The beta functions governing the scale-dependence of $G_k$ and $\Lambda_k$ are then obtained by restoring the extrinsic curvature terms in \eqref{hessian:flat} and evaluating the resulting operator traces via standard heat-kernel techniques. The result has been obtained in \cite{Biemans:2017zca} and is conveniently expressed in terms of the dimensionless couplings $\lambda_k \equiv \Lambda_k k^{-2}$ and $g_k = G_k k^{D-2}$. Restricting to $D=3+1$ dimensions it reads
\be\label{betafcts}
\p_t \lambda_k = \beta_\lambda(g,\lambda) \, , \quad \p_t g_k = \left(2 + \eta_N\right) g \, , 
\ee
where
\be
\beta_\lambda = \left(\eta_N - 2\right) \lambda - \tfrac{g}{24 \pi} \left(30 +3\eta_N - \tfrac{36-6 \eta_N}{1-2\lambda} - \tfrac{12- 2\eta_N}{2- 3 \lambda} \right) \, , 
\ee
and
\be
\eta_N = \frac{g B_1(\lambda)}{1-gB_2(\lambda)} \, . 
\ee
The functions $B_1$ and $B_2$ depend on $\lambda$ only and read
\be
\begin{split}
	B_1(\lambda)  = & \, \tfrac{1}{72 \pi} \left( -328 + \tfrac{156}{1-2\lambda}  + \tfrac{24}{2-3\lambda} - \tfrac{132}{(1-2\lambda)^2} + \tfrac{309}{(2-3\lambda)^2} \right) \, , \\
B_2(\lambda) = & \, - \tfrac{1}{144 \pi} \left( 4 + \tfrac{78}{1-2\lambda}  + \tfrac{12}{2-3\lambda} - \tfrac{44}{(1-2\lambda)^2} + \tfrac{103}{(2-3\lambda)^2} \right) \, . 
\end{split}
\ee

\begin{figure}[t!]
	\begin{center}
		\includegraphics[width=0.75\textwidth]{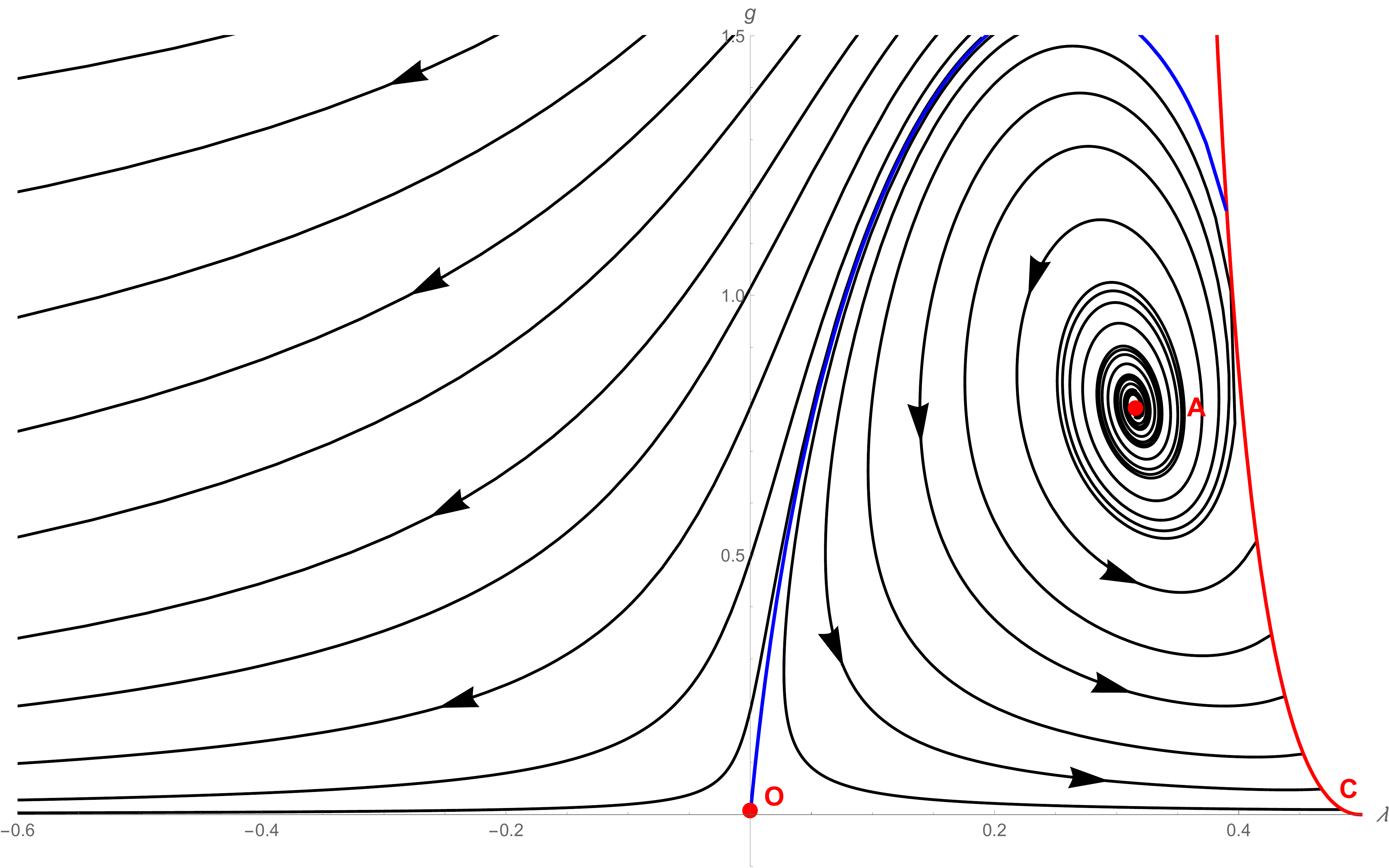}
	\end{center}
	\caption{\label{Fig:flows} Phase diagram obtained from integrating the beta functions \eqref{betafcts} numerically. The flow is governed by the interplay of the NGFP (labeled A) and the GFP (labeled O). The red line indicates a singular line of the flow where $\eta_N$ diverges. Arrows point towards lower values of $k$. (From \cite{Biemans:2017zca}.)}
\end{figure}
In the context of Asymptotic Safety, the most important feature of the beta functions are their fixed points where, by definition, the beta functions vanish simultaneously. In the present case, this entails $\beta_\lambda(g_*,\lambda_*) = 0, \beta_g(g_*,\lambda_*) = 0$. In the vicinity of a fixed point the properties of the RG flow can be studied by linearizing the beta functions. The linearized flow is then determined by the stability matrix ${\bf B}_{ij} \equiv \left. \frac{\p \beta_{g_i}}{\p g_j} \right|_{g = g_*}$. Defining the stability coefficients $\theta_i$ as minus the eigenvalues of ${\bf B}_{ij}$, eigendirections with Re$(\theta_i) > 0$ attract the flow as $k\rightarrow \infty$ while Re$(\theta_i) < 0$ implies that the corresponding eigendirection is UV-repulsive.

Restricting to the physically interesting region with positive Newton's coupling, $g \ge 0$, the system \eqref{betafcts} gives rise to two fixed points. The Gaussian fixed point (GFP) is located at the origin, $(\lambda_*, g_*) = (0,0)$, and its  stability coefficients are determined by the classical mass dimension of the couplings. In addition there is a non-Gaussian fixed point (NGFP) situated at
\be\label{NGFP1}
 g_* = 0.785 \, , \qquad \lambda_* =  0.315 \, , \qquad g_* \lambda_* = 0.248 \, . 
\ee
The NGFP comes with a complex pair of critical exponents, 
\be\label{critexp}
\theta_{1,2} = 0.503 \pm  5.377 i \, , 
\ee
 indicating that it  acts as a spiraling UV attractor for the RG trajectories in its vicinity. This is the same characteristic behavior of the NGFP found when evaluating the RG flow on foliated spacetimes using the Matsubara formalism \cite{Manrique:2011jc,Rechenberger:2012dt}, and a vast range of studies building on the covariant formalism, see \cite{Biemans:2016rvp} for a detailed comparison. The relations \eqref{NGFP1} entail that the dimensionful couplings at the fixed point scale according to
 \be\label{dimful}
 G_k = \frac{g_*}{k^2} \, , \qquad  \Lambda_k = \lambda_* \, k^2 \, . 
 \ee

The full phase diagram, shown in Fig.\ \ref{Fig:flows}, is obtained by integrating the system \eqref{betafcts} numerically. In the physical region the structure of the flow is determined by three features: a singular line associated with a divergence of the anomalous dimension $\eta_N$ (red line) bounds the region to the right. The flow to the left of this line is governed by the interplay of the NGFP (marked by A) and the GFP (marked by O). Flows in the vicinity of O are characterized by $g_k \ll 1$ and exhibit classical properties in the sense that the dimensionful couplings $G_k$ and $\Lambda_k$ are (approximately) $k$-independent in this regime. Depending on whether the trajectories flow to the left (right) of the blue separation line the value of $\Lambda_k$ in this classical regime is negative (positive). In \cite{Reuter:2001ag} these classes of solutions have been termed Type Ia and Type IIIa, respectively. The single trajectory ending at the GFP gives rise to a vanishing infrared value of the cosmological constant and is referred to as Type IIa. The diagram also illustrates that the NGFP has the appropriate features for providing the high-energy completion of the RG trajectories leaving the classical regime. A rather peculiar feature of the flow results from the interplay between the specific value of the critical exponents \eqref{critexp} and the vicinity of the fixed point to the singular line. In this particular configuration the singular line cuts through the RG trajectories undergoing a crossover from the NGFP to the classical regime. Since this feature is absent in the companion studies on foliated backgrounds \cite{Biemans:2016rvp,Houthoff:2017oam} and in the covariant computation \cite{Reuter:2001ag} we expect that this behavior will not persist in a refined approximation.

%-----------------------------------------------------------
\section{Self-consistent classical backgrounds}
\label{sect:4}
%-----------------------------------------------------------
A primary motivation for considering the FRGE in the presence of a foliation structure is its close relation to the Causal Dynamical Triangulations (CDT) program \cite{Ambjorn:2012jv}. The latter regularizes the gravitational partition sum by introducing piecewise linear building blocks and evaluates the resulting expression using Monte Carlo methods. The presence of a causal structure is thereby essential for obtaining structures resembling a macroscopic spacetime from these elementary building blocks \cite{Ambjorn:2004qm,Ambjorn:2005db}. Since the simulations are capable of tracking a finite number of building blocks only, CDT geometries have periodic boundary conditions in the time direction and are necessarily compact. Their topology is fixed to either $S^1 \times S^3$ or $S^1 \times T^3$ where $T^3$ denotes the flat three-torus.

A quantity that lends itself to a direct comparison in the FRG and CDT approach is the profile of the physical volumes $V_3(\tau)$ as a function of Euclidean time $\tau$. From the FRGE perspective, this information can be obtained by solving the equations of motion arising from by the effective average action $\Gamma_k$. Utilizing that the ansatz \eqref{GammaEH} is actually the Einstein-Hilbert action the equations determining the self-consistent backgrounds are Einstein's equations in the presence of a cosmological constant
\be\label{selfconsistent}
\left. \frac{\delta \Gamma_k[\hat{\chi}; \bar{\chi}]}{\delta \hat{\chi}} \right|_{\hat{\chi} = 0} = 0 \qquad \Longleftrightarrow \qquad R_{\mu\nu} - \tfrac{1}{2} g_{\mu\nu} R + g_{\mu\nu} \Lambda_k = 0 \, . 
\ee

\begin{figure}[t!]
	\includegraphics[width=0.45\textwidth]{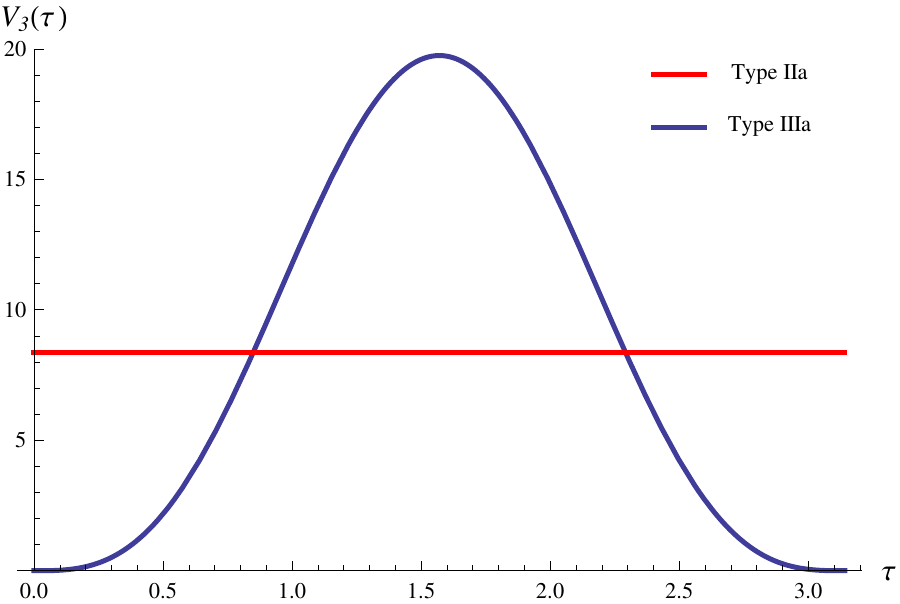} \;
	\includegraphics[width=0.45\textwidth]{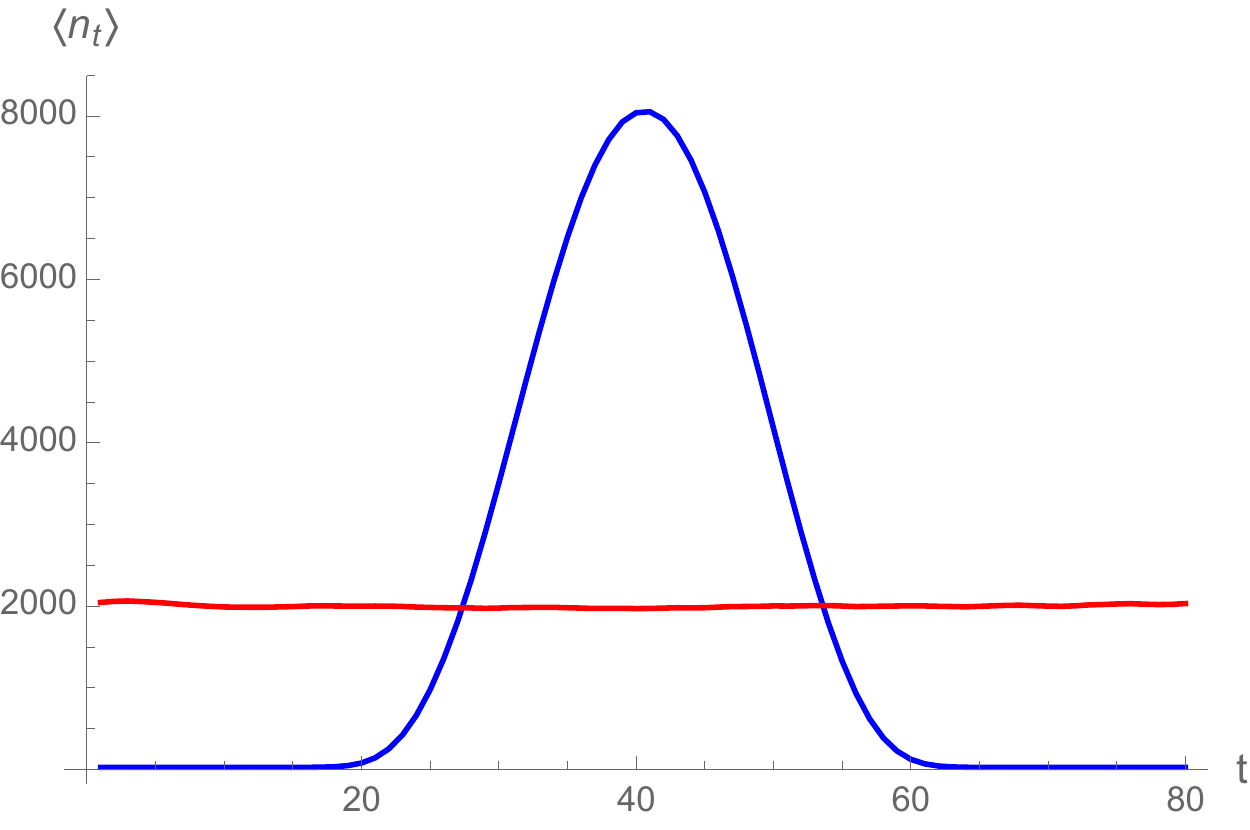} 
	\caption{\label{Fig:profiles} Volume profiles obtained from solving the self-consistency relations arising from the effective average action (left) and    CDT simulations \cite{Ambjorn:2011ph,Ambjorn:2016fbd,Ambjorn:2017ogo}  (right). The CDT results are obtained from simulations using topologies $S^1 \times T^3$ (red line) and $S^1 \times S^3$ (blue line) and show the expectation value $\langle n_t \rangle$ of the number of simplices found in a spatial slice. Their FRGE-counterparts result from evaluating the flow along trajectories of Type IIa (red line) and Type IIIa (blue line), respectively.}	
\end{figure}

We first focus on the solutions of this equation for RG trajectories of Type IIIa where $\Lambda_k > 0$ throughout. In this case, the solutions are given by four-spheres
\be\label{spheresol}
ds^2 = r^2 \, d\tau^2 + r^2 \, \sin^2(\tau) \left( d\psi^2 + \sin^2\psi \, d\theta^2 + \sin^2 \psi \, \sin^2 \theta \, d\phi^2 \right) \, ,
\ee
with radius $r^2 = \frac{3}{\Lambda_k}$. The time variable $\tau$ associated with the foliation takes values $\tau \in [0,\pi[$, mimicking the periodic boundary conditions encountered in the CDT framework. The profile for the spatial volumes $V_3(\tau)$ is obtained by integrating \eqref{spheresol} on the compact slices where $\tau$ is constant
\be\label{V3TypeIIIa}
\mbox{{Type IIIa}:} \qquad V_3(\tau) = 2 \pi^2 \, \left(\tfrac{3}{\Lambda_k} \right)^{3/2} \, \sin^3(\tau) \, . 
\ee
Thus $V_3(\tau)$ possesses the $\sin^3(\tau)$-profile characteristic for Euclidean de Sitter space. In the classical regime, where $\Lambda_k \simeq \Lambda_0$ is independent of $k$, $V_3(\tau)$ inherits this $k$-independence. The volume is then set by $\Lambda_0$ and can be used to identify the underlying RG trajectory. Typical volume profiles  obtained as a self-consistent solution of the effective average action in the classical regime and from the CDT program are shown as blue curves in Fig.\ \ref{Fig:profiles}, showing striking agreement.
In the NGFP regime the scaling relations \eqref{dimful} imply that $V_3(\tau) \propto k^{-3}$ vanishes as $k \rightarrow \infty$. This characteristic scaling behavior may be useful to identify the NGFP in CDT simulations.

The RG trajectory of Type IIa, connecting the NGFP to the GFP, plays a special role in the phase diagram shown in Fig.\ \ref{Fig:flows}. In this case the value of the cosmological constant vanishes in the infrared, $\Lambda_0 = 0$. The self-consistent (compact) solutions of \eqref{selfconsistent} then  have the topology $S^1 \times T^3$ and the spatial slices are given by three-tori with arbitrary, time-independent volume $V_3(\tau) = V_3$. 
This solution together with the volume profiles measured on CDT configurations with topology $S^1 \times T^3$ are shown as the red curves in Fig.\ \ref{Fig:profiles}. While in CDT the total volume of the configuration is fixed by the number of simplices the value $V_3$ determined from the effective average action is a free parameter which can be adjusted to match the CDT result. Following up on the comparison of spectral dimensions \cite{Reuter:2011ah}, the matching of the long-distance properties arising from the microscopic quantum descriptions of spacetime constitutes the next promising step in connecting the FRG and CDT frameworks in a systematic way.

%-----------------------------------------------------------------
\section{Summary of results and outlook}
\label{sect:5}
%-----------------------------------------------------------------
Initial works on functional renormalization group flows in the ADM formalism \cite{Manrique:2011jc,Rechenberger:2012dt,Contillo:2013fua} used a background $S^1 \times S^d$ and proper-time gauge, fixing the lapse function and shift vector to their background values. In this setting the radius of $S^d$ is taken to be time-independent so that the (Euclidean) time-direction associated with the $S^1$ constitutes a global Killing vector field. This feature allows to analytically continue the flow equation to Lorentzian signature. The fluctuations along the time direction can then be taken into account through Matsubara sums which, in contrast to the heat-kernel computations, can be carried out in both Euclidean and Lorentzian signature. This shows that the NGFP known from Euclidean signature computations persists for Lorentzian signature as well \cite{Manrique:2011jc}.

A more systematic investigation of the gravitational RG flows on backgrounds with topology $S^1 \times S^d$ has been performed in \cite{Houthoff:2017oam}. Here it was found that the spurious singularities plaguing the initial works arise due to the proper-time gauge and are removed once harmonic gauge, realizing background diffeomorphism invariance, is adopted. The systematic comparison between the fixed point structure obtained from the linear and exponential ADM splits revealed that both settings support a fixed point structure suitable for Asymptotic Safety. The critical exponents characterizing the NGFP in these two settings differ systematically in their values, supporting the expectation that the systems belong to two different gravitational universality classes. 

The analysis of RG flows on flat Friedman-Lema\^{\i}tre-Robertson-Walker (FLRW) backgrounds \cite{Biemans:2016rvp,Biemans:2017zca} supports this picture. The NGFP seen in the gravitational setting is largely independent of the topology of the background spacetime. Supplementing the gravitational sector by minimally coupled (non self-interacting) matter fields \cite{Biemans:2017zca} showed that many gravity-matter systems including the matter content of the standard model of particle physics actually possess NGFPs as well. Depending on the details of the matter sector the spiraling UV-attractor seen in the pure-gravity case is turned into a UV-attractor with real critical exponents. The overall picture obtained from the ADM framework thereby resembles to one seen within the covariant approach \cite{Percacci:2002ie,Percacci:2003jz,Dona:2013qba,Becker:2017tcx} if the same approximations and coarse graining schemes are employed. For a review of the state of the art in exploring asymptotically safe gravity-matter systems we refer to the proceedings article by A.\ Eichhorn \cite{Eichhorn:2017egq}.

The formulation of RG flows on cosmological backgrounds makes the ADM framework predestined for studying cosmological fluctuation spectra from first principles. In particular, the field decomposition \eqref{fielddec} formulates the FRGE in the field variables used in cosmic perturbation theory. Following the proposal \cite{Wetterich:2015gya}, this feature gives direct access to the two-point correlation functions. Based on these developments, we expect that Asymptotic Safety is capable of shedding light on fundamental questions related to the initial singularity and evolution of the very early universe based on first principles. Consequences of Asymptotic Safety on the formation of black holes are summarized in the proceedings of A.\ Platania \cite{BKPproceedings}.

%-----------------------------------------------------------

\begin{acknowledgements}
We thank the organizers of the workshop Lema\^{\i}tre for hospitality and setting up this very stimulating event. Furthermore, we thank A.\ Bonanno, G.\  Gionti, W.\ Houthoff, A.\ Kurov, K.\ Meissner, and M.\ Reuter for insightful discussions, A.\ G\"orlich for providing us with Fig.\ \ref{Fig:profiles}, and J.\ Biemans for participating in the original work. The research of F.~S. is supported by the Netherlands Organisation for Scientific Research (NWO) within the Foundation for Fundamental Research on Matter (FOM) grants 13PR3137 and 13VP12.
\end{acknowledgements}

%-----------------------------------------------------------


\begin{thebibliography}{99}

% -------------- Short list of selected reviews ------------

%\cite{Niedermaier:2006wt}
\bibitem{Niedermaier:2006wt}
M.~Niedermaier and M.~Reuter,
\emph{The Asymptotic Safety Scenario in Quantum Gravity},
Living Rev.\ Rel.\  {\bf 9}, 5, (2006).
%doi:10.12942/lrr-2006-5
%%CITATION = doi:10.12942/lrr-2006-5;%%
%285 citations counted in INSPIRE as of 03 May 2016

%\cite{Codello:2008vh}
\bibitem{Codello:2008vh}
A.~Codello, R.~Percacci and C.~Rahmede,
\emph{Investigating the Ultraviolet Properties of Gravity with a Wilsonian Renormalization Group Equation},
Annals Phys.\  {\bf 324}, 414, (2009),
%doi:10.1016/j.aop.2008.08.008
arXiv:0805.2909.
%%CITATION = doi:10.1016/j.aop.2008.08.008;%%
%272 citations counted in INSPIRE as of 13 Aug 2016

%\cite{Reuter:2012id}
\bibitem{Reuter:2012id}
M.~Reuter and F.~Saueressig,
\emph{Quantum Einstein Gravity},''
New J.\ Phys.,  {\bf 14}, 055022, (2012),
arXiv:1202.2274.
%%CITATION = ARXIV:1202.2274;%%
%121 citations counted in INSPIRE as of 07 Oct 2015

\bibitem{Roberto:book}
R.~Percacci,
\emph{An introduction to covariant quantum gravity and asymptotic safety}, World Scientific, Singapore (2017).

%\cite{Reuter:2001ag}
\bibitem{Reuter:2001ag}
M.~Reuter and F.~Saueressig,
\emph{Renormalization group flow of quantum gravity in the Einstein-Hilbert truncation},
Phys.\ Rev.\ D, {\bf 65}, 065016, (2002),
%doi:10.1103/PhysRevD.65.065016
hep-th/0110054.
%%CITATION = doi:10.1103/PhysRevD.65.065016;%%
%279 citations counted in INSPIRE as of 11 Aug 2017

%\cite{Reuter:2004nx}
\bibitem{Reuter:2004nx} 
M.~Reuter and H.~Weyer,
\emph{Quantum gravity at astrophysical distances?},
JCAP {\bf 12}, 001, (2004),
%doi:10.1088/1475-7516/2004/12/001
hep-th/0410119.
%%CITATION = doi:10.1088/1475-7516/2004/12/001;%%
%133 citations counted in INSPIRE as of 30 Sep 2017

% --------------- ADM review -------------------------------

%\cite{Gourgoulhon:2007ue}
\bibitem{Gourgoulhon:2007ue} 
E.~Gourgoulhon,
\emph{3+1 formalism and bases of numerical relativity},
gr-qc/0703035.
%%CITATION = GR-QC/0703035;%%
%195 citations counted in INSPIRE as of 30 Sep 2017


% --------------- CDT review -------------------------------

%\cite{Ambjorn:2012jv}
\bibitem{Ambjorn:2012jv} 
J.~Ambj{\o}rn, A.~G{\"o}rlich, J.~Jurkiewicz and R.~Loll,
\emph{Nonperturbative Quantum Gravity},
Phys.\ Rept.,  {\bf 519}, 127, (2012),
%doi:10.1016/j.physrep.2012.03.007
arXiv:1203.3591.
%%CITATION = doi:10.1016/j.physrep.2012.03.007;%%
%155 citations counted in INSPIRE as of 28 Sep 2017


% ----------------------------------------------------------

%\cite{Wetterich:1992yh}
\bibitem{Wetterich:1992yh}
C.~Wetterich,
\emph{Exact evolution equation for the effective potential},
Phys.\ Lett.\ B, {\bf 301}, 90, (1993).
%doi:10.1016/0370-2693(93)90726-X
%%CITATION = doi:10.1016/0370-2693(93)90726-X;%%
%1098 citations counted in INSPIRE as of 13 Aug 2017

%\cite{Morris:1993qb}
\bibitem{Morris:1993qb}
T.~R.~Morris,
\emph{The Exact renormalization group and approximate solutions},
Int.\ J.\ Mod.\ Phys.\ A, {\bf 9}, 2411, (1994),
%  doi:10.1142/S0217751X94000972
hep-ph/9308265.
%%CITATION = doi:10.1142/S0217751X94000972;%%
%418 citations counted in INSPIRE as of 12 Aug 2017

%\cite{Reuter:1993kw}
\bibitem{Reuter:1993kw}
M.~Reuter and C.~Wetterich,
\emph{Effective average action for gauge theories and exact evolution equations},
Nucl.\ Phys.\ B, {\bf 417}, 181, (1994).
% doi:10.1016/0550-3213(94)90543-6
%%CITATION = doi:10.1016/0550-3213(94)90543-6;%%
%291 citations counted in INSPIRE as of 04 Sep 2017

%\cite{Reuter:1996cp}
\bibitem{Reuter:1996cp} 
M.~Reuter,
\emph{Nonperturbative evolution equation for quantum gravity},
Phys.\ Rev.\ D {\bf 57}, 971, (1998),
%doi:10.1103/PhysRevD.57.971
hep-th/9605030.
%%CITATION = doi:10.1103/PhysRevD.57.971;%%
%576 citations counted in INSPIRE as of 29 Sep 2017

%\cite{Nink:2015lmq}
\bibitem{Nink:2015lmq} 
A.~Nink and M.~Reuter,
\emph{The unitary conformal field theory behind 2D Asymptotic Safety},
JHEP, {\bf 02}, 167, (2016),
%doi:10.1007/JHEP02(2016)167
arXiv:1512.06805.
%%CITATION = doi:10.1007/JHEP02(2016)167;%%
%9 citations counted in INSPIRE as of 03 Oct 2017


%\cite{Biemans:2016rvp}
\bibitem{Biemans:2016rvp}
J.~Biemans, A.~Platania and F.~Saueressig,
\emph{Quantum gravity on foliated spacetimes: Asymptotically Safe and sound},
Phys.\ Rev.\ D, {\bf 95}, 086013, (2017), 
%doi:10.1103/PhysRevD.95.086013
arXiv:1609.04813.
%%CITATION = doi:10.1103/PhysRevD.95.086013;%%
%13 citations counted in INSPIRE as of 28 Sep 2017

%\cite{Biemans:2017zca}
\bibitem{Biemans:2017zca}
J.~Biemans, A.~Platania and F.~Saueressig,
\emph{Renormalization group fixed points of foliated gravity-matter systems},
JHEP, {\bf 05}, 093, (2017).
%doi:10.1007/JHEP05(2017)093
%%CITATION = doi:10.1007/JHEP05(2017)093;%%
%10 citations counted in INSPIRE as of 26 Sep 2017

% --------------------  

%\cite{Manrique:2011jc}
\bibitem{Manrique:2011jc}
E.~Manrique, S.~Rechenberger and F.~Saueressig,
\emph{Asymptotically Safe Lorentzian Gravity},
Phys.\ Rev.\ Lett.,  {\bf 106}, 251302, (2011),
%doi:10.1103/PhysRevLett.106.251302
arXiv:1102.5012.
%%CITATION = doi:10.1103/PhysRevLett.106.251302;%%
%57 citations counted in INSPIRE as of 28 Sep 2017

%\cite{Rechenberger:2012dt}
\bibitem{Rechenberger:2012dt}
S.~Rechenberger and F.~Saueressig,
\emph{A functional renormalization group equation for foliated spacetimes},
JHEP, {\bf 03}, 010, (2013), 
%doi:10.1007/JHEP03(2013)010
arXiv:1212.5114.
%%CITATION = doi:10.1007/JHEP03(2013)010;%%
%30 citations counted in INSPIRE as of 28 Sep 2017

%\cite{Contillo:2013fua}
\bibitem{Contillo:2013fua}
A.~Contillo, S.~Rechenberger and F.~Saueressig,
\emph{Renormalization group flow of Hořava-Lifshitz gravity at low energies},
JHEP,  {\bf 13}, 017, (2013),
%doi:10.1007/JHEP12(2013)017
arXiv:1309.7273.
%%CITATION = doi:10.1007/JHEP12(2013)017;%%
%18 citations counted in INSPIRE as of 28 Sep 2017

%\cite{Houthoff:2017oam}
\bibitem{Houthoff:2017oam}
W.~B.~Houthoff, A.~Kurov and F.~Saueressig,
\emph{Impact of topology in foliated Quantum Einstein Gravity},
Eur.\ Phys.\ J.\ C, {\bf 77}, 491, (2017),
%doi:10.1140/epjc/s10052-017-5046-8
arXiv:1705.01848.
%%CITATION = doi:10.1140/epjc/s10052-017-5046-8;%%
%4 citations counted in INSPIRE as of 28 Sep 2017

% ------------ discussion of covariant gauges --------------


%\cite{Ohta:2016npm}
\bibitem{Ohta:2016npm} 
N.~Ohta, R.~Percacci and A.~D.~Pereira,
\emph{Gauges and functional measures in quantum gravity I: Einstein theory},
JHEP, {\bf 06}, 115, (2016),
%doi:10.1007/JHEP06(2016)115
arXiv:1605.00454.
%%CITATION = doi:10.1007/JHEP06(2016)115;%%
%17 citations counted in INSPIRE as of 29 Sep 2017

%\cite{Ohta:2016jvw}
\bibitem{Ohta:2016jvw} 
N.~Ohta, R.~Percacci and A.~D.~Pereira,
\emph{Gauges and functional measures in quantum gravity II: Higher derivative gravity},
Eur.\ Phys.\ J.\ C, {\bf 77}, 611, (2017),
%doi:10.1140/epjc/s10052-017-5176-z
arXiv:1610.07991.
%%CITATION = doi:10.1140/epjc/s10052-017-5176-z;%%
%10 citations counted in INSPIRE as of 29 Sep 2017

% ------------ CDT profile works ---------------------------
%\cite{Ambjorn:2004qm}
\bibitem{Ambjorn:2004qm} 
J.~Ambjorn, J.~Jurkiewicz and R.~Loll,
\emph{Emergence of a 4-D world from causal quantum gravity},
Phys.\ Rev.\ Lett.,  {\bf 93}, 131301, (2004),
% doi:10.1103/PhysRevLett.93.131301
hep-th/0404156.
%%CITATION = doi:10.1103/PhysRevLett.93.131301;%%
%251 citations counted in INSPIRE as of 02 Oct 2017

%\cite{Ambjorn:2005db}
\bibitem{Ambjorn:2005db} 
J.~Ambjorn, J.~Jurkiewicz and R.~Loll,
\emph{Spectral dimension of the universe},
Phys.\ Rev.\ Lett.,  {\bf 95}, 171301, (2005),
%doi:10.1103/PhysRevLett.95.171301
hep-th/0505113.
%%CITATION = doi:10.1103/PhysRevLett.95.171301;%%
%328 citations counted in INSPIRE as of 02 Oct 2017


%\cite{Ambjorn:2011ph}
\bibitem{Ambjorn:2011ph} 
J.~Ambj{\o}rn, A.~G{\"o}rlich, J.~Jurkiewicz, R.~Loll, J.~Gizbert-Studnicki and T.~Trzesniewski,
\emph{The Semiclassical Limit of Causal Dynamical Triangulations},
Nucl.\ Phys.\ B, {\bf 849}, 144, (2011),
%doi:10.1016/j.nuclphysb.2011.03.019
arXiv:1102.3929.
%%CITATION = doi:10.1016/j.nuclphysb.2011.03.019;%%
%49 citations counted in INSPIRE as of 29 Sep 2017

%\cite{Ambjorn:2016fbd}
\bibitem{Ambjorn:2016fbd} 
J.~Ambj{\o}rn, Z.~Drogosz, J.~Gizbert-Studnicki, A.~G{\"o}rlich, J.~Jurkiewicz and D.~Nemeth,
\emph{Impact of topology in causal dynamical triangulations quantum gravity},
Phys.\ Rev.\ D, {\bf 94}, 044010, (2016),
%doi:10.1103/PhysRevD.94.044010
arXiv:1604.08786.
%%CITATION = doi:10.1103/PhysRevD.94.044010;%%
%3 citations counted in INSPIRE as of 29 Sep 2017

%\cite{Ambjorn:2017ogo}
\bibitem{Ambjorn:2017ogo} 
J.~Ambj{\o}rn, J.~Gizbert-Studnicki, A.~G{\"o}rlich, K.~Grosvenor and J.~Jurkiewicz,
\emph{Four-dimensional CDT with toroidal topology},''
Nucl.\ Phys.\ B, {\bf 922}, 226, (2017),
%doi:10.1016/j.nuclphysb.2017.06.026
arXiv:1705.07653.
%%CITATION = doi:10.1016/j.nuclphysb.2017.06.026;%%

%\cite{Reuter:2011ah}
\bibitem{Reuter:2011ah} 
M.~Reuter and F.~Saueressig,
\emph{Fractal space-times under the microscope: A Renormalization Group view on Monte Carlo data},
JHEP, {\bf 12}, 012, (2011),
%doi:10.1007/JHEP12(2011)012
arXiv:1110.5224.
%%CITATION = doi:10.1007/JHEP12(2011)012;%%
%62 citations counted in INSPIRE as of 03 Oct 2017

% ------------ matter field works --------------------------

%\cite{Percacci:2002ie}
\bibitem{Percacci:2002ie}
R.~Percacci and D.~Perini,
\emph{Constraints on matter from asymptotic safety},
Phys.\ Rev.\ D, {\bf 67}, 081503, (2003),
%doi:10.1103/PhysRevD.67.081503
hep-th/0207033.
%%CITATION = doi:10.1103/PhysRevD.67.081503;%%
%166 citations counted in INSPIRE as of 28 Sep 2017

%\cite{Percacci:2003jz}
\bibitem{Percacci:2003jz}
R.~Percacci and D.~Perini,
\emph{Asymptotic safety of gravity coupled to matter},
Phys.\ Rev.\ D, {\bf 68}, 044018, (2003),
%doi:10.1103/PhysRevD.68.044018
hep-th/0304222.
%%CITATION = doi:10.1103/PhysRevD.68.044018;%%
%191 citations counted in INSPIRE as of 28 Sep 2017

%\cite{Dona:2013qba}
\bibitem{Dona:2013qba}
P.~Donà, A.~Eichhorn and R.~Percacci,
\emph{Matter matters in asymptotically safe quantum gravity},
Phys.\ Rev.\ D, {\bf 89}, 084035, (2014),
%doi:10.1103/PhysRevD.89.084035
arXiv:1311.2898.
%%CITATION = doi:10.1103/PhysRevD.89.084035;%%
%78 citations counted in INSPIRE as of 28 Sep 2017

%\cite{Becker:2017tcx}
\bibitem{Becker:2017tcx}
D.~Becker, C.~Ripken and F.~Saueressig,
\emph{On avoiding Ostrogradski instabilities within Asymptotic Safety},
arXiv:1709.09098 [hep-th].
%%CITATION = ARXIV:1709.09098;%%

%\cite{Eichhorn:2017egq}
\bibitem{Eichhorn:2017egq}
A.~Eichhorn,
\emph{Status of the asymptotic safety paradigm for quantum gravity and matter}, 
proceedings of the workshop Lema\^{\i}tre, arXiv:1709.03696 [gr-qc].
%%CITATION = ARXIV:1709.03696;%%
%1 citations counted in INSPIRE as of 28 Sep 2017

%\cite{Wetterich:2015gya}
\bibitem{Wetterich:2015gya} 
C.~Wetterich,
\emph{Cosmic fluctuations from a quantum effective action},
Phys.\ Rev.\ D, {\bf 92}, 083507, (2015),
%doi:10.1103/PhysRevD.92.083507
arXiv:1503.07860.
%%CITATION = doi:10.1103/PhysRevD.92.083507;%%
%8 citations counted in INSPIRE as of 28 Sep 2017





\bibitem{BKPproceedings}
A.~Bonanno, B.~Koch and A.~Platania,
\emph{Gravitational collapse in Quantum Einstein Gravity},
 proceedings of the workshop Lema\^{\i}tre, to appear.


% etc
\end{thebibliography}
\end{document}